\documentclass[11pt]{article}
\usepackage{color}
\title{\huge
	{Sample Size for Concurrent Species\\ \underline{Detection in a Species-Rich Assemblage}}}
\usepackage{multirow}

\usepackage{tabularx}
\usepackage{amsmath, amsthm, amssymb, mathrsfs}
\usepackage{bm}
\usepackage{xcolor-patch}

\date{}
\begin{document}
	\maketitle
	\begin{flushleft}
		\author{\textbf{Ali T. Haidar$^1$, Abbas Al-Hakim$^2$, and Zhiyi Zhang$^3$}\newline\\$^1$Department of Geography, Faculty of Humanities, First Section, Beirut, Lebanon. Corresponding author. E-mail address: altalhdr@gmail.com\newline\\$^2$Department of Mathematics, American University of Beirut (AUB), Beirut, Lebanon.\newline\\$^3$Department of Mathematics and Statistics, University of N. Carolina at Charlotte (UNCC),\\ N. Carolina, USA.
		}
	\end{flushleft}
	\section*{KEYWORDS}
	Alpha, binomial distribution, biodiversity, biostratigraphy, biozone, calcareous nannofossils, confidence level, species count, diversity, ecology, evolution, geography, geologic time, geology, independent and identically distributed sample, law of inclusion and exclusion, marker species, microfossils, micropaleontology, minimum sample size, multinomial distribution, multiple species detection, oceanography, paleoceanography, paleoecology, percentage, probability of failure, sampling effort, sediment, species proportion, relative abundance, stratigraphic correlation, stratigraphic markers, stratigraphy, subset, taxon.
	\section*{ABSTRACT}
	Monitoring the distribution of microfossils in stratigraphic successions is an essential tool for biostratigraphic, evolutionary and paleoecologic/paleoceanographic studies. To estimate the relative abundance ($\%$) of a given species, it is necessary to estimate in advance the minimum number of specimens to be used in the count (n). This requires an \textit{a priori} assumption about a specified level of confidence, and about the species population proportion ($p$). It is common use to apply the binomial distribution to determine n to detect the presence of \textit{more} than one species in the same sample, although the multinomial distribution should necessarily be used instead.
	The mathematical theory of sample size computation using the multinomial distribution is adapted to the computation of n for any number of species to be detected together (K) at any level of confidence. Easy-to-use extensive tables show n, for a combination of K and p. These tables indicate a large difference for n between that indicated by the binomial and those by the multinomial distribution when many species are to be detected simultaneously. Counting only 300 specimens (with 95$\%$ confidence level) or 500 (99$\%$) is not enough to detect more than one taxon.
	The reconstructed history of the micro-biosphere may therefore, in many instances, need to be largely revised. This revision should affect our understanding of the ecological and evolutionary relationships between the past changes in the biosphere and the other major reservoirs (hydrosphere, geosphere and atmosphere). In biostratigraphy and biochronology, using a much larger sample size, when more than one marker species is to be detected in the neighborhood of the same biozone boundary, may help clarifying the nature of the apparent inconsistencies given by the observed reversals in the ordinal (rank) biostratigraphic data shown as intersections of the correlation lines.
	\section{GEOLOGICAL AND ECOLOGICAL BACKGROUND}
	\subsection{Introduction}
	Statistical techniques are used to compensate for sampling effects (for a review, see Hayek and Buzas, 1997; Moore \textit{et al.}, 2007). The relative abundances of species (or taxa in general) and/or their remains based on sample counts are used to estimate their original proportion in the living communities and the fossil assemblages, hence the necessity to compute n to be used in the estimate of the abundance of a given species at a specified level of confidence. Obviously, the higher the number of species to be contemporaneously detected in a sample, the larger the required value of n, implying a larger sampling effort. Determining precisely this minimum sample size is particularly relevant in ecological, evolutionary and stratigraphic studies. For the detection of the position of a stratigraphic event, micropaleontologists analyze a short geologic (stratigraphic) time interval rather than a single point in geologic time. In other words, biostratigraphic events are detected over a number of samples in the stratigraphic succession, and more than one marker species may appear (or disappear) during the time interval represented by these samples, thus the recurrent need for monitoring precisely the relative abundance of all of these marker species (usually attempting to detect them together) in the same sample, at a given levels of confidence. Hereafter we use ``together`` to indicate that the species were detected using the same count rather than by repeating the count to detect another species in the same sample.
	\subsection{Current Practice to Determine Sample Size}
	\subsubsection*{$\textit{1.2.1}$ \textit{Binomial Distribution}}
	Micropaleontologists are usually interested in detecting co-occurring species mainly at 3 confidence levels (90, 95 and 99$\%$). It may seem contradictory (circular thinking) to assume a certain population proportion for the determination of n, if this sample will in turn be used to estimate this (originally given) population proportion (see this reasoning in Moore \textit{et al.}, 2007). However, since the original definition of most biostratigraphic events (definition of a biozone boundary) requires \textit{a priori} a change through time of the species proportion in the population so that it becomes either larger (or smaller) than a pre-determined value (\textit{e.g.}, an event is identified when the proportion of a given stratigraphic marker exceeds 5$\%$, see Backman \textit{et al.}, 2012 as an example) at a given stratigraphic level (\textit{e.g.}, first or last appearance, \textit{etc.}), then there is a necessity to determine n using the species proportion (\textit{i.e.}, searching this size in a table by choosing the assumed population proportion) as indicated by the original biostratigraphic definition of the event in order to check for the possible presence of such an event. In other words, there is a need to perform the taxonomic analysis by knowing in advance how many specimens to count in any sample. To estimate the sample size (n) for given species abundances, found in a species-rich assemblage, both binomial (\textit{e.g.}, Dennison and Hay, 1967; Fatela and Taborda, 2002) and multinomial (\textit{e.g.}, Moore \textit{et al.}, 2007) distributions are used. The use of the binomial distribution model is based on the assumption that the underlying population is a homogeneous random mixture (Agterberg, 1990). Using binomial distributions, the required n is estimated in order to detect a given species chosen \textit{a priori}, present in the population at a certain proportion, with a given degree of confidence (1 - probability of failure). The probability of failure (fail to detect a species in the sample even when this is present in the population = incorrectly rejecting the null hypothesis) is denoted with P, whereas the species proportion in the population is denoted with p. A standard procedure in micropaleontology was therefore to count 299 specimens per sample - usually used as 300 specimens - for the confidence level of 95$\%$, in order to detect a species present in the population at 1$\%$ (or 459 specimens with 99$\%$ confidence - usually used as 500 specimens). Note that these values of n, extracted from the graph of Dennison and Hay (1967), are approximate. In the sedimentary record, to detect the presence of a given taxon, a minimum relative abundance of it should be observed. In other words, it is not enough to observe only one specimen of this taxon to confirm its presence. This is due to a variety of factors (see discussion in Agterberg, 1990), including for example either the (random) vertical mixing of pelagic sediments (Guinasso and Schink, 1975), or its discrete vertical shuffling (Haidar, 2015).
	\subsubsection*{$\textit{1.2.2}$ \textit {\small The Paradox Given by the Use of the Multinomial Distribution is only Apparent}}
	How could a count be valid for the detection of a given species alone, and the same count be valid for the detection of another given species alone, but the \textit{same} count cannot be used for the detection of both species together? For two biostratigraphers looking at the results of a count at the same time, it may seem that one of them would have the ``right`` to interpret the results as valid for the detection of the first species, whereas the other would interpret the same results for the detection of the second species. This paradox is however only apparent, and the use of the multinomial distribution implies that the same count cannot be valid for the detection of both species together. Indeed, it is possible to clarify how using the same count is misleading by applying an (opposite) example on half of the count. Two biostratigraphers looking, at the same time, at the results of only half of a count (counting only the half of the total number of individuals required following the binomial distribution - to detect only one species). One of them would think s/he has the ``right`` to interpret the results as representing the first half of the count, whereas the other would interpret the same results as valid for the second half. Clearly, counting only half of the required total number of individuals is not enough to detect the presence of a species.
	\subsubsection*{$\textit{1.2.3}$ \textit {\small Binomial Distribution Leads to a Down-sized Sample}}
	Values of n were commonly and erroneously applied, by relying on a binomial distribution, regardless of how many species were to be detected together. Computing n, using Dennison and Hay (1967), to detect concurrent (multiple) species (together) leads obviously to a down-sized sample. In biodiversity studies, this error in the estimate of n implies the detection of any species at a confidence level lower than that required. In biostratigraphy, species are most of the time rare around their times of origination and extinction. A down-sized sample may therefore lead to an error in the determination of the position of a biostratigraphic event. In chronostratigraphy, this error would show, on average, an apparent lowest (highest) occurrence of a stratigraphic marker later (earlier) than what it could be detected using an appropriate n. This is may be, in part, the reason for which an apparent inconsistency is (commonly) observed as reversals in the ranking of biostratigraphic events. This change of the time-order relationships of the biostratigraphic events between different stratigraphic successions appears in the form of intersections of the correlation lines (see examples in Hills and Thierstein, 1989).
	\subsubsection*{$\textit{1.2.4}$\textit { Previously Available Multinomial Sample Sizes Are Neither Applicable to Biostratigraphy Nor to Biodiversity}}
	In a multinomial distribution (see \textit{e.g.}, Degroot and Schervish, 2012), several categories or species are considered with proportions p$_{1}$, ...; p$_{k}$ , where K is the number of categories. Moore \textit{et al}. (2007) used a multinomial model to compute n necessary to detect a species at a given confidence level. However, they required to estimate all the relative species abundances (p$_{1}$, ...; p$_{k}$) at a fixed confidence level when a degree of similarity d is predetermined. This degree of similarity is not the species proportion in the population, but rather a predetermined margin of error identical for the estimate of all the species proportions. As this is not the major concern in biostratigraphy, our approach is rather similar to that used by Dennison and Hay (1967), in terms of setting in advance both the original species proportion and the confidence level, but applied to a multinomial (rather than to a binomial) distribution, by seeking the n needed to detect K species in the same sample (\textit{i.e.}, simultaneously).
	\subsection{Sample Size Based on a Multinomial Distribution}
	\subsubsection*{$\textit{1.3.1}$ \textit {Practical Estimate of Sample Size}}
	This paper uses a binomial distribution only if K = 1 (only one pre-determined species to be detected in a sample). The distribution becomes multinomial for K $>$ 1, with more than one pre-determined species to be detected together, and with the original individual proportions p$_{1}$, p$_{2}$, ..., p$_{k}$, p$_{k+1}$ all pre-determined. The multinomial model computes n to detect all of the species of interest together, rather than any (particular subset) of them, each of them being present at a possibly different relative abundance in the population. The multinomial distribution gives precise values of n. Although the full mathematical derivation is provided (see below), this does not need to be understood in order to apply the method, as detailed and easy-to-use statistical tables are also provided. Due to space limitation, our tables cannot provide values of n for every possible combination of confidence level, species proportion in the population, and number of concurrent species of interest. The theoretical background provided compensates for the missed computation of any possible combination of these parameters. The computation of n can be done by solving the inequality provided (see below) using standard mathematical software Mathematica$^{\tiny\textregistered}$, Matlab$^{\tiny\textregistered}$, Maple$^{\tiny\textregistered}$, or the open source Maxima$^{\tiny\textregistered}$ (descendent of Macsyma$^{\tiny\textregistered}$), \textit{etc}. In other words, there is no need to solve it by hand in any practical situation. Furthermore, if one needs to skip the computation, and still get a useful estimate of n for a combination of the above-mentioned parameters that is not available in any table, it is recommended to use a sample size (value) available in a table corresponding to a slightly more confidence (\textit{i.e.}, choosing a confidence level slightly larger than that required or a slightly smaller value of $\alpha$), to a slightly smaller species proportion (a slightly smaller value of $p$ than that required), and/or to a slightly larger number of concurrent species of interest to be detected simultaneously (a slightly larger value of K). Although this procedure will have the disadvantage of giving sample sizes (slightly) larger than those that would be precisely calculated (larger counting effort), it will have the advantage of providing an additional gain in the accuracy of the estimate of the species proportion (conservative estimate).
	\subsubsection*{$\textit{1.3.2}$ \textit{\small Example and Practice}}
	Persico \textit{et al.} (2012) counted at least 500 specimens per sample. In one stratigraphic interval (at around 40 mbsf) of an investigated succession (ODP Hole 738B, Southern Kerguelen Plateau in the Southern Indian Ocean), they attempted to detect 4 species (\textit{C. reticulatum}, \textit{C. eoaltus}, \textit{I. recurvus}, and \textit{R. oamaruensis}) together (their Fig. 2). It is may be recommended to consider that the neighborhood of the biozone boundary - the stratigraphic interval useful for the detection of the lowest (or highest) occurrence of a species - is not only the indicated depth, but expanded to cover also the stratigraphic distance including at least a couple of samples lying immediately below (a couple above) the \textit{position} of this occurrence.\\\hspace{2em}
	
	It is worth noting that the species were anyway detected in the stratigraphic succession even using the binomial distribution. The focus here is rather on whether the bionomial distribution allowed for the detection of the lowest (common) or the highest occurrence for each of the 4 investigated species in the proper stratigraphic position at the required level of confidence.\\\hspace{2em}
	
	Based on the previous way of thinking that was leading to the use of the binomial distribution, a count of 500 (it should actually be of only 459) specimens should have been enough to detect \textit{any} of these species present in the population at 1$\%$ or more, with 99$\%$ confidence. However, based on the interpretation of the multinomial distribution provided in this paper, and in order to detect \textit{all} of the 4 species \textit{together}, each of them being present at 1$\%$ and with 99$\%$ confidence, at least 599 specimens (Table 4, column 2, row 3, with K = 4, $p_{1}$ = $p_{2}$ = $p_{3}$ = $p_{4}$ = 0.01, and probability of failure = 0.01) should have been counted in every sample of this stratigraphic interval. Based on Table 4, a count of only 500 specimens in these 3 samples would still be enough to detect these species, but at 95$\%$ confidence (column 2, row 2, with K = 4, $p_{1}$ = $p_{2}$ = $p_{3}$ = $p_{4}$ = 0.01, and probability of failure = 0.05 gives a minimum of 437 specimens).\\\hspace{2em}
	
	Persico \textit{et al.} (2012) were indirectly able to largely overcome this problem (of not counting the extra needed 599 - 500 = 99 specimens per sample) by additionally scanning 2 long traverses (80 mm$^2$). However, by counting traverses, it is not possible to quantitatively assess neither the minimum percentage of the detected species in the population, nor the degree of confidence reached during the analysis.\\\hspace{2em}
	
	In another stratigraphic interval (at around 100 mbsf), Persico \textit{et al.} (2012) monitored together the lowest occurrence of \textit{C. reticulatum} and the highest occurrence of \textit{R. clatrata}. According to the multinomial distribution, a minimum of 527 specimens per sample (Table 2 (with K = 2), column 3, row 14, $p_{1}$ = $p_{2}$ = 0.01, and probability of failure = 0.01) is required to detect both species together, each of them being present in the population at 1$\%$, and with 99$\%$ confidence level. A count of only 500 specimens in these 3 samples (the sample where the species is detected, and the adjacent 2 samples where the species is not detected in the count) would theoretically be enough to detect both of these 2 species at 99$\%$ confidence, when one of these 2 species is present at 1$\%$ (or at a higher proportion) in the population, but this count will allow for the detection of the remaining species \textit{only if} this is present at 5$\%$ (or at a higher proportion) in the population (Table 2, column 3, row 9, $p_{1}$ = 0.01, $p_{2}$ = 0.05, and probability of failure = 0.01, gives a sample size of 459 specimens). Conversely, a count of only 500 specimens in these 3 samples would be theoretically enough, when both of the species are present at 1$\%$ (or at a higher proportion) in the population, to detect both of the species only at 95$\%$ confidence (Table 2, column 2, row 14, $p_{1}$ = $p_{2}$ = 0.01, and probability of failure = 0.05, gives a sample size of 366 specimens).\\\hspace{2em}
	
	When searching in the Tables 2 and 3, if the required proportion $p_{1}$, $p_{2}$ or $p_{3}$ of the first species is not found in the first column of the appropriate table, then the user should try to find the proportion of this first species in the second column (if not, then in the third, and so on). To avoid redundancy, Tables 2 and 3 contain all the possible combinations for the selected p and n, but not in any possible order. Obviously, this order doesn't influence the computation of the sample size.
	\subsubsection*{$\textit{1.3.3}$ \textit{Count Method}}
	The procedure of species count using the multinomial distribution could be made in two steps. Since, prior to the microscopic investigations, the biostratigrapher cannot know the sample stratgraphic position with respect to the standard biozonations, it is not clear \textit{a priori} which definition of a biostratigraphic event is to be used (\textit{i.e.}, not clear how many are the predetermined stratigraphic markers to investigate together, and at which proportions these marker species need to be detected). A preliminary count should therefore be made, using a sample size based on a binomial distribution. This is equivalent to trying to detect only one species (K = 1) per count. The results of this preliminary count are used to approximately identify the position of the stratigraphic event. Only then, it would be possible to make a preliminary identification of all the concurrent marker species possibly present at the given stratigraphic interval. After this preliminary identification, the correction of the sample size could be made based on the multinomial distribution.\\\hspace{2em}
	
	The number of the marker species (K) is not determined for a single sample, but rather for a short geologic time interval, corresponding to the neighborhood of the biozone boundary. This stratigraphic interval might be covered by few samples (not only one), with K that applies to \textit{each} sample in this interval. To determine K for each sample, there is a need to detect the total number of marker species present in the whole stratigraphic interval, rather than in the specific single sample.\\\hspace{2em}
	
	Consider for example a biostratigraphic event (a lowest occurrence of a marker species) that is usually identified (by going upward in the stratigraphic succession) when its relative abundance exceeds 1$\%$. If, during the identification of this stratigraphic event, the count based on a sample size relative to a binomial distribution revealed the presence of one additional marker species in the same stratigraphic interval, then an additional count shoud be made on the remaining individuals of the same sample to increase n from that based on a binomial distribution into that based on a multinomial distribution having K = 2. If, using the binomial distribution, 2 additional marker species were detected during the count (instead of only 1), then the remaining needed part of the count should increase the sample size so that the total number of species counted would correspond to that given by a multinomial distribution with K = 3. In other words, the additional counts should be made using the remaining part of the sample, with the difference between binomial and multinomial sample size to be added to those of the previous count that was based on a sample size relative to the binomial distribution. This is to reach a total sample size computed according to that given by the multinomial distribution, \textit{after} the definite value of K has been determined.\\\hspace{2em}
	
	Only in rare cases, it would be necessary to repeat this procedure. This would be equivalent to making more than one count on (the remaining individuals of) the same sample. This repetition practically leads to a repetitive increase of the value of n (\textit{i.e.}, increase of n more than once). This is necessary when more than one additional count reveals an increase the in number of marker species (of interest). In this case, the total number of species revealed with all of the counts together (original count based on a binomial distribution, and all of the additional counts based on multinomial distributions) would have to be used to select the appropriate multinomial table and compute n accordingly.\\\hspace{2em}
	
	A similar count procedure could be used while estimating the diversity of an assemblage, although in some diversity studies there could be more interest in time slices rather than in stratigraphic intervals. A preliminary count could be made, based on a binomial distribution, to roughly estimate the number of species to be detected simultaneously (K), and their corresponding approximate proportion in the population (p$_{n}$), before determining precise values of n based on the multinomial distribution.\\\hspace{2em}
	
	The number of times an additional count must be made is quite limited, despite the fact that the continuous increase of n gives more chance to reveal the presence of additional rare species of interest (increase of K). This is because, in stratigraphy for example, in most cases, the extremely rare species are usually not used in the standard biozonations.\\\hspace{2em}
	
	Using this count method for the multinomial distribution, it becomes easy to correct both the position of stratigraphic events and the diversity estimates previously established using counts based on the binomial distribution. Indeed, once the correction of the sample size is revealed to be necessary, it would be easier to rely on the results of any previous count (rather than to repeat the count on any sample from the beginning), and only add to these results those of a new count using a newly prepared part of the same sample in order to reach a value of n suitable according the multinomial distribution.
	\section{STATISTICAL FOUNDATIONS}
	Consider a population with multiple species and their
	proportions $\{p_{k}^{*};k\geq 1\}$, where $k=1,\cdots$, is an index
	for species and there could be infinitely many species in the
	population. Suppose a micropaleontologist would like to take an
	independent and identically distributed sample of size $n$ to
	include $K\geq 1$ pre-determined markers simultaneously with
	probability $1-\alpha$. How large should the sample size be?\\\hspace{2em}
	
	Say that a micropaleontologist or a biodiversity specialist considers $K$ marker species, then the
	rest of the species can be lumped into one group. Consequently, we
	have a multinomial distribution $\{p_{k};k=1,\cdots,K,K+1\}$ with
	$K$ parameters, $p_{k}=p_{k}^{*}$, for $k=1,\cdots,K$, and
	$p_{K+1}=\sum_{k\geq K+1}p_{k}^{*}$. Clearly
	$\sum_{k=1}^{K+1}p_{k}=\sum_{k\geq 1}p_{k}^{*}=1$. We want for some
	small $\alpha \in (0,1)$
	\[\begin{array}{l}
	P\left(Y_{1}\geq 1 \mbox{ $\&$ } Y_{2}\geq 1 \mbox{ $\&$ }\cdots \mbox{ $\&$ } Y_{K}\geq 1\right)\geq 1-\alpha;
	\end{array}\]\\\hspace{2em}
	
	where $Y_{k}$ is the observed sample frequency of the $k^{th}$ marker in the sample, or
	equivalently by using the complement and then applying the law of
	inclusion and exclusion
	\begin{equation}
	\begin{array}{l}
	\alpha\geq P\left(Y_{1}=0 \mbox{ or } Y_{2}=0 \mbox{ or }\cdots \mbox{ or } Y_{K}=0\right) \\ \\
	=\sum_{k=1}^{K}P\left(Y_{k}=0\right) \\ \\
	+(-1)^{2-1}\sum_{1\leq k_{1}<k_{2}\leq K}P\left(Y_{k_{1}}=0\mbox{ $\&$ } Y_{k_{2}}=0\right) \\ \\
	+(-1)^{3-1}\sum_{1\leq k_{1}<k_{2}<k_{3}\leq K}P\left(Y_{k_{1}}=0
	\mbox{ $\&$ }Y_{k_{2}}=0\mbox{ $\&$ }Y_{k_{3}}=0\right) \\ \\
	+\cdots \\ \\
	+(-1)^{K-1}P\left(Y_{1}=0 \mbox{ $\&$ }Y_{2}=0\mbox{ $\&$ }\cdots \mbox{ $\&$ }Y_{K}=0\right).
	\label{ex1}
	\end{array}
	\end{equation}\\\hspace{2em}
	
	Provided $\{p_{k}\}$ and $\alpha$ are known, the last
	expression in (\ref{ex1}) is a function of the sample size $n$,
	which therefore can be in principle solved in that inequality, {\it
		\textit{i.e.}}, the smallest integer value of $n$ satisfying
	(\ref{ex1}). However, when $K$ is large, solving that inequality
	could be a bit tedious by hand. Zooming in on any additive term of
	the last expression of (\ref{ex1}) and ignoring the sign, by
	symmetry we have the term involving $m$ ($m\leq K$) indices,
	$k_{1},\cdots,k_{m}$,
	\[\begin{array}{l}
	\sum_{1\leq k_{1}<k_{2}< \cdots <k_{m}\leq K} P\left(Y_{k_{1}}\mbox{ $\&$ }\cdots\mbox{ $\&$ }Y_{k_{m}}=0\right)=\sum (1-p_{k_{1}}-p_{k_{2}}-\cdots -p_{k_{m}})^{n}
	\end{array}\]\\\hspace{2em}
	
	where the second $\sum$ is over the same index set as the first $\sum$. Therefore the general form of the inequality in (\ref{ex1}) is
	\begin{equation}
	(a_{1}^{n}+a_{2}^{n}+\cdots+a_{I}^{n}) -(b_{1}^{n}+b_{2}^{n}+\cdots+b_{J}^{n})\leq \alpha
	\label{ex2}
	\end{equation}\\\hspace{2em}
	
	where $I$ and $J$ are some (possibly very large) positive
	integers. The objective is to find the smallest integer value $n$
	satisfying the inequality in (\ref{ex2}).\\\hspace{2em}
	
	Since in the practice of micropaleontology, and in some biodiversity studies, it is most common that, for an
	anticipated sample, one would only work with a few particular
	marker species, we will first reduce the general problem to a few special cases
	and solve the inequality for K=1, 2, and 3.
	Then we will proceed with the calculation for K= 4 or more only under specific assumptions.
	\subsection{$K=1$.}
	For illustration purpose, let us work out the case of $K=1$. In this
	case, (\ref{ex1}) becomes
	\[
	\begin{array}{l}
	P\left(Y_{1}=0\right)=(1-p_{1})^{n}\leq \alpha.
	\end{array}
	\] or equivalently
	\begin{equation}
	\begin{array}{l}
	n=\left\{
	\begin{array}{ll}
	\left\lfloor \frac{\ln(\alpha)}{\ln(1-p_{1})}\right\rfloor,& \mbox{if $\left\lfloor \frac{\ln(\alpha)}{\ln(1-p_{1})}\right\rfloor$ is a positive integer}, \\ & \\
	\left\lfloor \frac{\ln(\alpha)}{\ln(1-p_{1})}\right\rfloor+1,&\mbox{otherwise,}
	\end{array}
	\right.
	\end{array}
	\label{ex3}
	\end{equation}
	where $\lfloor \cdot \rfloor$ is the floor of a
	real number. The formula in (\ref{ex3}) gives the precise value of
	the minimum sample size required to cover at least $K=1$ specimen in
	a sample with probability $\alpha$. The figure of Dennison and Hay (1967) is essentially an approximation of this simple
	formula. Let us write $p_{1}=p$. Table 1 below gives the required
	sample size for various values of $p$, and various values of $1-\alpha$. 
	\begin{table*}[ht]
		{Table 1. Minimum sample sizes required for detecting $K=1$ marker species.}
		\[
		\begin{array}{|c|c|c|c|c|c|} \hline
		1-\alpha & p=0.001 & p=0.005  & p=0.010 & p=0.050 & p=0.100 \\ \hline
		0.90 & 2302&460&230&45&22 \\
		0.95 &2995&598&299&59&29 \\
		0.99 &4603&919&459&90&44 \\ \hline
		\end{array}
		\]	
	\end{table*}\\
	The readers may wish to compare Table~1 to the values extracted from the figure of Dennison and Hay (1967). Note that the commonly used value $n=500$ is highly approximate as this should be $n=459$.
	\subsection{$K=2$.} When $K=2$, based on (\ref{ex1}) we want to find the minimum integer value of $n$ such that
	\begin{equation}
	\begin{array}{c}
	(1-p_{1})^{n}+(1-p_{2})^{n}- (1-p_{1}-p_{2})^{n} \leq \alpha.
	\end{array}
	\label{ex4}
	\end{equation}
	The formula in (\ref{ex4}) is for any set of values
	of $p_{1}\in (0,1)$ and $p_{2}\in (0,1)$ subject to $p_{1}+p_{2}\in
	(0,1)$. For illustration, we produce the Table~2 for various values
	of $p_{1}$ and $p_{2}$.
	\subsection{$K=3$.} When $K=3$, based on (\ref{ex1}) we want to find the minimum integer value of $n$ such that
	\begin{equation}
	\begin{array}{l}
	(1-p_{1})^{n}+(1-p_{2})^{n}+(1-p_{3})^{n} \\ \\
	- (1-p_{1}-p_{2})^{n} - (1-p_{1}-p_{3})^{n}- (1-p_{2}-p_{3})^{n} \\ \\
	+ (1-p_{1}-p_{2}-p_{3})^n  \leq \alpha.
	\end{array}
	\label{ex5}
	\end{equation}
	The formula in (\ref{ex5}) is for any set of values
	of $p_{1}\in (0,1)$, $p_{2}\in (0,1)$ and $p_{3}\in (0,1)$ subject
	to $p_{1}+p_{2}+p_{3}\in (0,1)$. For illustration, we produce the
	Table~3 for various values of $p_{1}$, $p_{2}$, and $p_{3}$.
	\subsection{$K=4$ or more.} Table~4 indicates the minimum integer value of $n$ when many species are to be detected together. The use of this table requires that all of the species should be present at the same original proportion in the population. This table is also essential in biodiversity studies. It is based on a formula that applies properly only for large values of K. For small values of K, the formula in (\ref{ex6}) becomes approximate (the formula in (\ref{ex7}) is used to calculate table 4). The error due to the approximation given by this table, when present, is conservative (always an overestimate) of the sample size. In other words, by using Table 4, the biostratigrapher is in reality counting only very few specimens more than required by the minimum smple size necessary.\\ \\
	\begin{align*}
	&\sum_{i=0}^{K}(^K_i)(-1)^i(1-ip)^n\gtrsim1-\alpha\\
	&\sum_{i=0}^{K}(^K_i)(-1)^i ({1}-\frac{i}{1/p})^\frac{np}{p}\gtrsim1-\alpha\\
	&\sum_{i=0}^{K}(^K_i)(-e{^{-np}})^i\gtrsim1-\alpha\\ 
	&(1-e{^{-np}})^K\gtrsim1-\alpha\\ 
	&1-e{^{-np}}\gtrsim(1-\alpha)^\frac{1}{K}\\ 
	&e{^{-np}}\lesssim1-(1-\alpha)^\frac{1}{K}\\ 
	&-np\lesssim\ln\lfloor1-(1-\alpha)^\frac{1}{K}\rfloor
	\end{align*}
	\begin{align}
	&n\gtrsim\frac{1}{p}\ln\lceil\frac{1}{1-(1-\alpha)^\frac{1}{K}}\rceil
	\label{ex6} \\ 
	&n\approx\lfloor\frac{1}{p}\ln(\frac{1}{1-(1-\alpha)^\frac{1}{K}})\rfloor-2\label{ex7}
	\end{align}
	\begin{table*}[ht!]
		{Table 2. Minimum sample sizes required for detecting K = 2 marker species together showing some possible combinations of $\alpha$ (0.1, 0.05, 0.01, 0.005, 0.001, 0.0005, and 0.0001) with $p$ (0.1, 0.05, 0.01, 0.005, 0.001, 0.0005, and 0.0001). Example: 598 specimens (individuals) are to be counted (see row 4, column 2, header rows and header columns not included) in order to detect 2 species together (using the same count), with one of the species being present at 0.5$\%$ (0.005) in the population and the other at 10$\%$ (0.1), with a confidence level of 95$\%$ ($\alpha$ = 0.05).}
		\begin{center}
			\[
			\begin{array}{rr|rrrrrrr}	 & \multicolumn{5}{r}{\huge\bm\alpha} \\ \cline{3-9}
			\multicolumn{1}{c}{\textbf{\textit{p}1}} & \multicolumn{1}{c}{\textbf{\textit{p}2}} & \multicolumn{1}{|c}{\textbf{0.1}} & \multicolumn{1}{c}{\textbf{0.05}} & \multicolumn{1}{c}{\textbf{0.01}} & \multicolumn{1}{c}{\textbf{0.005}} & \multicolumn{1}{c}{\textbf{0.001}} & \multicolumn{1}{c}{\textbf{0.0005}} & \multicolumn{1}{c}{\textbf{0.0001}} \\
			\hline
			0.1   & 0.1   & 29    & 35    & 51    & 57    & 73    & 79    & 94 \\
			0.05  & 0.1   & 47    & 60    & 90    & 104   & 135   & 149   & 180 \\
			0.01  & 0.1   & 230   & 299   & 459   & 528   & 688   & 757   & 917 \\
			0.005 & 0.1   & 460   & 598   & 919   & 1058  & 1379  & 1517  & 1838 \\
			0.001 & 0.1   & 2302  & 2995  & 4603  & 5296  & 6905  & 7598  & 9206 \\
			0.0005 & 0.1   & 4605  & 5990  & 9209  & 10594 & 13813 & 15199 & 18417 \\
			0.0001 & 0.1   & 23025 & 29956 & 46050 & 52981 & 69075 & 76006 & 92099 \\
			0.05  & 0.05  & 58    & 72    & 104   & 117   & 149   & 162   & 194 \\
			0.01  & 0.05  & 230   & 299   & 459   & 528   & 688   & 757   & 917 \\
			0.005 & 0.05  & 460   & 598   & 919   & 1058  & 1379  & 1517  & 1838 \\
			0.001 & 0.05  & 2302  & 2995  & 4603  & 5296  & 6905  & 7598  & 9206 \\
			0.0005 & 0.05  & 4605  & 5990  & 9209  & 10594 & 13813 & 15199 & 18417 \\
			0.0001 & 0.05  & 23025 & 29956 & 46050 & 52981 & 69075 & 76006 & 92099 \\
			0.01  & 0.01  & 296   & 366   & 527   & 597   & 757   & 826   & 986 \\
			0.005 & 0.01  & 476   & 607   & 921   & 1058  & 1379  & 1517  & 1838 \\
			0.001 & 0.01  & 2302  & 2995  & 4603  & 5296  & 6905  & 7598  & 9206 \\
			0.0005 & 0.01  & 4605  & 5990  & 9209  & 10594 & 13813 & 15199 & 18417 \\
			0.0001 & 0.01  & 23025 & 29956 & 46050 & 52981 & 69075 & 76006 & 92099 \\
			0.005 & 0.005 & 593   & 734   & 1057  & 1196  & 1517  & 1655  & 1976 \\
			0.001 & 0.005 & 2302  & 2995  & 4603  & 5296  & 6905  & 7598  & 9206 \\
			0.0005 & 0.005 & 4605  & 5990  & 9209  & 10594 & 13813 & 15199 & 18417 \\
			0.0001 & 0.005 & 23025 & 29956 & 46050 & 52981 & 69075 & 76006 & 92099 \\
			0.001 & 0.001 & 2969  & 3675  & 5294  & 5988  & 7597  & 8290  & 9899 \\
			0.0005 & 0.001 & 4765  & 6079  & 9228  & 10604 & 13815 & 15199 & 18417 \\
			0.0001 & 0.001 & 23025 & 29956 & 46050 & 52981 & 69075 & 76006 & 92099 \\
			0.0005 & 0.0005 & 5939  & 7351  & 10589 & 11978 & 15198 & 16584 & 19802 \\
			0.0001 & 0.0005 & 23026 & 29956 & 46050 & 52981 & 69075 & 76006 & 92099 \\
			0.0001 & 0.0001 & 29696 & 36760 & 52956 & 59900 & 76003 & 82936 & 99030 \\
			\end{array}
			\]\\	
		\end{center}	
	\end{table*}
	\newpage
	\begin{table*}[ht!]
		{Table 3. Minimum sample sizes required for detecting K = 3 marker species together showing some possible combinations of $\alpha$ (0.1, 0.05, 0.01, 0.005, 0.001, 0.0005, and 0.0001) with $p$ (0.1, 0.05, 0.01, 0.005, 0.001, 0.0005, and 0.0001). Example: 4603 specimens (individuals) are to be counted (see row 11, column 3, header rows and header columns not included) in order to detect 3 species together (using the same count), with one species being present at 0.1$\%$ (0.001) in the population, another species at 5$\%$ (0.05), and the remaining  at 10$\%$ (0.1), with a confidence level of 99$\%$ ($\alpha$ = 0.01).}	
		\[
		\begin{array}{rrr|rrrrrrr}
		& & \multicolumn{5}{r}{\huge\bm\alpha} \\ \cline{4-10}
		\textbf{\textit{p}1} & \textbf{\textit{p}2} & \textbf{\textit{p}3} & \textbf{0.1} & \textbf{0.05} & \textbf{0.01} & \textbf{0.005} & \textbf{0.001} & \textbf{0.0005} & \textbf{0.0001} \\
		\hline
		0.1   & 0.1   & 0.1   & 33    & 39    & 55    & 61    & 76    & 83    & 98 \\
		0.05  & 0.1   & 0.1   & 48    & 60    & 91    & 104   & 135   & 149   & 180 \\
		0.01  & 0.1   & 0.1   & 230   & 299   & 459   & 528   & 688   & 757   & 917 \\
		0.005 & 0.1   & 0.1   & 460   & 598   & 919   & 1058  & 1379  & 1517  & 1838 \\
		0.001 & 0.1   & 0.1   & 2302  & 2995  & 4603  & 5296  & 6905  & 7598  & 9206 \\
		0.0005 & 0.1   & 0.1   & 4605  & 5990  & 9209  & 10594 & 13813 & 15199 & 18417 \\
		0.0001 & 0.1   & 0.1   & 23025 & 29956 & 46050 & 52981 & 69075 & 76006 & 92099 \\
		0.05  & 0.05  & 0.1   & 59    & 72    & 104   & 117   & 149   & 162   & 194 \\
		0.01  & 0.05  & 0.1   & 230   & 299   & 459   & 528   & 688   & 757   & 917 \\
		0.005 & 0.05  & 0.1   & 460   & 598   & 919   & 1058  & 1379  & 1517  & 1838 \\
		0.001 & 0.05  & 0.1   & 2302  & 2995  & 4603  & 5296  & 6905  & 7598  & 9206 \\
		0.0005 & 0.05  & 0.1   & 4605  & 5990  & 9209  & 10594 & 13813 & 15199 & 18417 \\
		0.0001 & 0.05  & 0.1   & 23025 & 29956 & 46050 & 52981 & 69075 & 76006 & 92099 \\
		0.01  & 0.01  & 0.1   & 296   & 366   & 527   & 597   & 757   & 826   & 986 \\
		0.005 & 0.01  & 0.1   & 476   & 607   & 921   & 1058  & 1379  & 1517  & 1838 \\
		0.001 & 0.01  & 0.1   & 2302  & 2995  & 4603  & 5296  & 6905  & 7598  & 9206 \\
		0.0005 & 0.01  & 0.1   & 4605  & 5990  & 9209  & 10594 & 13813 & 15199 & 18417 \\
		0.0001 & 0.01  & 0.1   & 23025 & 29956 & 46050 & 52981 & 69075 & 76006 & 92099 \\
		0.005 & 0.005 & 0.1   & 593   & 734   & 1057  & 1196  & 1517  & 1655  & 1976 \\
		0.001 & 0.005 & 0.1   & 2302  & 2995  & 4603  & 5296  & 6905  & 7598  & 9206 \\
		0.0005 & 0.005 & 0.1   & 4605  & 5990  & 9209  & 10594 & 13813 & 15199 & 18417 \\
		0.0001 & 0.005 & 0.1   & 23025 & 29956 & 46050 & 52981 & 69075 & 76006 & 92099 \\
		0.001 & 0.001 & 0.1   & 2969  & 3675  & 5294  & 5988  & 7597  & 8290  & 9899 \\
		0.0005 & 0.001 & 0.1   & 4765  & 6079  & 9228  & 10604 & 13815 & 15199 & 18417 \\
		0.0001 & 0.001 & 0.1   & 23025 & 29956 & 46050 & 52981 & 69075 & 76006 & 92099 \\
		0.0005 & 0.0005 & 0.1   & 5939  & 7351  & 10589 & 11978 & 15198 & 16584 & 19802 \\
		0.0001 & 0.0005 & 0.1   & 23026 & 29956 & 46050 & 52981 & 69075 & 76006 & 92099 \\
		0.0001 & 0.0001 & 0.1   & 29696 & 36760 & 52956 & 59900 & 76003 & 82936 & 99030 \\
		0.05  & 0.05  & 0.05  & 66    & 80    & 112   & 125   & 157   & 170   & 201 \\
		0.01  & 0.05  & 0.05  & 230   & 299   & 459   & 528   & 688   & 757   & 917 \\
		0.005 & 0.05  & 0.05  & 460   & 598   & 919   & 1058  & 1379  & 1517  & 1838 \\
		0.001 & 0.05  & 0.05  & 2302  & 2995  & 4603  & 5296  & 6905  & 7598  & 9206 \\
		0.0005 & 0.05  & 0.05  & 4605  & 5990  & 9209  & 10594 & 13813 & 15199 & 18417 \\
		0.0001 & 0.05  & 0.05  & 23025 & 29956 & 46050 & 52981 & 69075 & 76006 & 92099 \\
		0.01  & 0.01  & 0.05  & 296   & 366   & 527   & 597   & 757   & 826   & 986 \\
		0.005 & 0.01  & 0.05  & 476   & 607   & 921   & 1058  & 1379  & 1517  & 1838 \\
		0.001 & 0.01  & 0.05  & 2302  & 2995  & 4603  & 5296  & 6905  & 7598  & 9206 \\
		0.0005 & 0.01  & 0.05  & 4605  & 5990  & 9209  & 10594 & 13813 & 15199 & 18417 \\
		\end{array}
		\]
		
	\end{table*}
	\begin{table}[t]
		\[
		\begin{array}{rrr|rrrrrrr}
		\textbf{\textit{p}1} & \textbf{\textit{p}2} & \textbf{\textit{p}3} & \textbf{0.1} & \textbf{0.05} & \textbf{0.01} & \textbf{0.005} & \textbf{0.001} & \textbf{0.0005} & \textbf{0.0001} \\
		\hline
		0.0001 & 0.01  & 0.05  & 23025 & 29956 & 46050 & 52981 & 69075 & 76006 & 92099 \\
		0.005 & 0.005 & 0.05  & 593   & 734   & 1057  & 1196  & 1517  & 1655  & 1976 \\
		0.001 & 0.005 & 0.05  & 2302  & 2995  & 4603  & 5296  & 6905  & 7598  & 9206 \\	
		0.0005 & 0.005 & 0.05  & 4605  & 5990  & 9209  & 10594 & 13813 & 15199 & 18417 \\
		0.0001 & 0.005 & 0.05  & 23025 & 29956 & 46050 & 52981 & 69075 & 76006 & 92099 \\
		0.001 & 0.001 & 0.05  & 2969  & 3675  & 5294  & 5988  & 7597  & 8290  & 9899 \\
		0.0005 & 0.001 & 0.05  & 4765  & 6079  & 9228  & 10604 & 13815 & 15199 & 18417 \\
		0.0001 & 0.001 & 0.05  & 23025 & 29956 & 46050 & 52981 & 69075 & 76006 & 92099 \\	
		
		0.0005 & 0.0005 & 0.05  & 5939  & 7351  & 10589 & 11978 & 15198 & 16584 & 19802 \\
		0.0001 & 0.0005 & 0.05  & 23026 & 29956 & 46050 & 52981 & 69075 & 76006 & 92099 \\
		0.0001 & 0.0001 & 0.05  & 29696 & 36760 & 52956 & 59900 & 76003 & 82936 & 99030 \\
		0.01  & 0.01  & 0.01  & 336   & 406   & 568   & 637   & 797   & 866   & 1026 \\
		0.005 & 0.01  & 0.01  & 489   & 615   & 923   & 1059  & 1379  & 1517  & 1838 \\
		0.001 & 0.01  & 0.01  & 2302  & 2995  & 4603  & 5296  & 6905  & 7598  & 9206 \\
		0.0005 & 0.01  & 0.01  & 4605  & 5990  & 9209  & 10594 & 13813 & 15199 & 18417 \\
		0.0001 & 0.01  & 0.01  & 23025 & 29956 & 46050 & 52981 & 69075 & 76006 & 92099 \\
		0.005 & 0.005 & 0.01  & 598   & 736   & 1058  & 1196  & 1517  & 1655  & 1976 \\
		0.001 & 0.005 & 0.01  & 2302  & 2995  & 4603  & 5296  & 6905  & 7598  & 9206 \\
		0.0005 & 0.005 & 0.01  & 4605  & 5990  & 9209  & 10594 & 13813 & 15199 & 18417 \\
		0.0001 & 0.005 & 0.01  & 23025 & 29956 & 46050 & 52981 & 69075 & 76006 & 92099 \\
		0.001 & 0.001 & 0.01  & 2969  & 3675  & 5294  & 5988  & 7597  & 8290  & 9899 \\
		0.0005 & 0.001 & 0.01  & 4765  & 6079  & 9228  & 10604 & 13815 & 15199 & 18417 \\
		0.0001 & 0.001 & 0.01  & 23025 & 29956 & 46050 & 52981 & 69075 & 76006 & 92099 \\
		0.0005 & 0.0005 & 0.01  & 5939  & 7351  & 10589 & 11978 & 15198 & 16584 & 19802 \\
		0.0001 & 0.0005 & 0.01  & 23026 & 29956 & 46050 & 52981 & 69075 & 76006 & 92099 \\
		0.0001 & 0.0001 & 0.01  & 29696 & 36760 & 52956 & 59900 & 76003 & 82936 & 99030 \\
		0.005 & 0.005 & 0.005 & 672   & 814   & 1138  & 1276  & 1598  & 1736  & 2057 \\
		0.001 & 0.005 & 0.005 & 2302  & 2995  & 4603  & 5296  & 6905  & 7598  & 9206 \\
		0.0005 & 0.005 & 0.005 & 4605  & 5990  & 9209  & 10594 & 13813 & 15199 & 18417 \\
		0.0001 & 0.005 & 0.005 & 23025 & 29956 & 46050 & 52981 & 69075 & 76006 & 92099 \\
		0.001 & 0.001 & 0.005 & 2969  & 3675  & 5294  & 5988  & 7597  & 8290  & 9899 \\
		0.0005 & 0.001 & 0.005 & 4765  & 6079  & 9228  & 10604 & 13815 & 15199 & 18417 \\
		0.0001 & 0.001 & 0.005 & 23025 & 29956 & 46050 & 52981 & 69075 & 76006 & 92099 \\
		0.0005 & 0.0005 & 0.005 & 5939  & 7351  & 10589 & 11978 & 15198 & 16584 & 19802 \\
		0.0001 & 0.0005 & 0.005 & 23026 & 29956 & 46050 & 52981 & 69075 & 76006 & 92099 \\
		0.0001 & 0.0001 & 0.005 & 29696 & 36760 & 52956 & 59900 & 76003 & 82936 & 99030 \\
		0.001 & 0.001 & 0.001 & 3365  & 4076  & 5698  & 6393  & 8003  & 8695  & 10304 \\
		0.0005 & 0.001 & 0.001 & 4896  & 6158  & 9247  & 10614 & 13817 & 15200 & 18417 \\
		0.0001 & 0.001 & 0.001 & 23025 & 29956 & 46050 & 52981 & 69075 & 76006 & 92099 \\
		0.0005 & 0.0005 & 0.001 & 5986  & 7375  & 10594 & 11980 & 15199 & 16584 & 19803 \\
		0.0001 & 0.0005 & 0.001 & 23026 & 29956 & 46050 & 52981 & 69075 & 76006 & 92099 \\
		0.0001 & 0.0001 & 0.001 & 29696 & 36760 & 52956 & 59900 & 76003 & 82936 & 99030 \\
		0.0005 & 0.0005 & 0.0005 & 6732  & 8153  & 11399 & 12788 & 16009 & 17395 & 20613 \\
		0.0001 & 0.0005 & 0.0005 & 23027 & 29956 & 46050 & 52981 & 69075 & 76006 & 92099 \\
		0.0001 & 0.0001 & 0.0005 & 29697 & 36760 & 52956 & 59900 & 76003 & 82936 & 99030 \\
		0.0001 & 0.0001 & 0.0001 & 33664 & 40772 & 57002 & 63950 & 80057 & 86990 & 100001
		\end{array}
		\]	
	\end{table}
	\begin{table*}[!t]
		{Table 4. Minimum sample sizes required to detect K = 4, 5, ..., 30 marker species together (using the same count), at the confidence levels of 90$\%$ (probability of failure = $\alpha$ = 0.1), 95$\%$ ($\alpha$ = 0.05), 99$\%$ ($\alpha$ = 0.01), and 99.9$\%$ ($\alpha$ = 0.001). This table shows the sample size when this original proportion is either 5$\%$ (0.05), or 1$\%$ (0.01), or 0.5$\%$ (0.005), or 0.1$\%$ (0.001) or 0.05$\%$ (0.0005), or 0.01$\%$ (0.0001). Example: 477 (instead of only 300) specimens (individuals) are to be counted (see row 10, column 2, header rows and header columns not included) in order to detect 6 species together, with each of these species being present at 1$\%$ (0.01) in the population, with a confidence level of 95$\%$ ($\alpha$ = 0.05). Note that the error in this case would be significant (actually huge), as more than 40$\%$ of the specimens would be missed in the count if the binomial distribution was applied. }
		\begin{center}
			\[
			\begin{array}{rr|rrrrrr}	&  \multicolumn{5}{r}{\textit{\textbf{p}}} \\ \cline{3-8}
			\textbf{K} & {\huge\bm\alpha} & \textbf{ 0.05} & \textbf{0.01} & \textbf{ 0.005} & \textbf{0.001} & \textbf{ 0.0005} & \textbf{0.0001} \\
			\hline
			4     & 0.1   & 73    & 365   & 730   & 3650  & 7300  & 36499 \\
			4     & 0.05  & 88    & 437   & 873   & 4363  & 8726  & 43629 \\
			4     & 0.01  & 120   & 599   & 1198  & 5988  & 11976 & 59877 \\
			4     & 0.001 & 166   & 830   & 1659  & 8294  & 16588 & 82937 \\
			5     & 0.1   & 78    & 388   & 775   & 3871  & 7741  & 38704 \\
			5     & 0.05  & 92    & 459   & 917   & 4585  & 9170  & 45848 \\
			5     & 0.01  & 125   & 622   & 1243  & 6211  & 12422 & 62106 \\
			5     & 0.001 & 171   & 852   & 1704  & 8517  & 17034 & 85168 \\
			6     & 0.1   & 82    & 406   & 811   & 4051  & 8102  & 40509 \\
			6     & 0.05  & 96    & 477   & 954   & 4767  & 9533  & 47663 \\
			6     & 0.01  & 128   & 640   & 1279  & 6393  & 12786 & 63928 \\
			6     & 0.001 & 174   & 870   & 1740  & 8700  & 17399 & 86991 \\
			7     & 0.1   & 85    & 421   & 841   & 4204  & 8408  & 42038 \\
			7     & 0.05  & 99    & 492   & 984   & 4920  & 9840  & 49198 \\
			7     & 0.01  & 131   & 655   & 1310  & 6547  & 13094 & 65468 \\
			7     & 0.001 & 178   & 886   & 1771  & 8854  & 17707 & 88533 \\
			8     & 0.1   & 87    & 434   & 868   & 4337  & 8673  & 43364 \\
			8     & 0.05  & 102   & 506   & 1011  & 5053  & 10106 & 50529 \\
			8     & 0.01  & 134   & 669   & 1337  & 6681  & 13361 & 66803 \\
			8     & 0.001 & 180   & 899   & 1798  & 8987  & 17974 & 89868 \\
			9     & 0.1   & 90    & 446   & 891   & 4454  & 8907  & 44535 \\
			9     & 0.05  & 104   & 518   & 1035  & 5171  & 10341 & 51703 \\
			9     & 0.01  & 136   & 680   & 1360  & 6798  & 13596 & 67980 \\
			9     & 0.001 & 183   & 911   & 1821  & 9105  & 18210 & 91046 \\
			10    & 0.1   & 92    & 456   & 912   & 4559  & 9117  & 45583 \\
			10    & 0.05  & 106   & 528   & 1056  & 5276  & 10551 & 52754 \\
			10    & 0.01  & 139   & 691   & 1381  & 6904  & 13807 & 69033 \\
			10    & 0.001 & 185   & 921   & 1842  & 9210  & 18420 & 92099 \\
			11    & 0.1   & 94    & 466   & 931   & 4654  & 9307  & 46531 \\
			11    & 0.05  & 108   & 538   & 1075  & 5371  & 10741 & 53705 \\
			
			\end{array}
			\]\\
		\end{center}
	\end{table*}%
	\newpage
	\begin{table}[htbp]
		\begin{center}
			\[
			\begin{array}{rr|rrrrrr}
			\textbf{K} & {\huge\bm\alpha} & \textbf{ 0.05} & \textbf{ 0.01} & \textbf{0.005} & \textbf{ 0.001} & \textbf{ 0.0005} & \textbf{ 0.0001} \\
			\hline
			11    & 0.01  & 140   & 700   & 1400  & 6999  & 13998 & 69986 \\
			11    & 0.001 & 187   & 931   & 1862  & 9306  & 18611 & 93052 \\
			12    & 0.1   & 95    & 474   & 948   & 4740  & 9480  & 47397 \\
			12    & 0.05  & 110   & 546   & 1092  & 5458  & 10915 & 54573 \\
			12    & 0.01  & 142   & 709   & 1418  & 7086  & 14171 & 70855 \\
			12    & 0.001 & 188   & 940   & 1879  & 9393  & 18785 & 93923 \\
			13    & 0.1   & 97    & 482   & 964   & 4820  & 9639  & 48194 \\		
			13    & 0.05  & 111   & 554   & 1108  & 5538  & 11075 & 55372 \\
			13    & 0.01  & 144   & 717   & 1434  & 7166  & 14331 & 71655 \\
			13    & 0.001 & 190   & 948   & 1895  & 9473  & 18945 & 94723 \\
			14    & 0.1   & 98    & 490   & 979   & 4894  & 9787  & 48932 \\
			14    & 0.05  & 113   & 562   & 1123  & 5612  & 11223 & 56111 \\
			14    & 0.01  & 145   & 724   & 1448  & 7240  & 14480 & 72396 \\
			14    & 0.001 & 191   & 955   & 1910  & 9547  & 19093 & 95464 \\
			15    & 0.1   & 100   & 497   & 993   & 4962  & 9924  & 49620 \\
			15    & 0.05  & 114   & 568   & 1136  & 5680  & 11360 & 56800 \\
			15    & 0.01  & 147   & 731   & 1462  & 7309  & 14618 & 73086 \\
			15    & 0.001 & 193   & 962   & 1924  & 9616  & 19231 & 96154 \\
			16    & 0.1   & 101   & 503   & 1006  & 5027  & 10053 & 50263 \\
			16    & 0.05  & 115   & 575   & 1149  & 5745  & 11489 & 57444 \\
			16    & 0.01  & 148   & 738   & 1475  & 7374  & 14747 & 73731 \\
			16    & 0.001 & 194   & 968   & 1936  & 9680  & 19360 & 96799 \\
			17    & 0.1   & 102   & 509   & 1018  & 5087  & 10174 & 50867 \\
			17    & 0.05  & 117   & 581   & 1161  & 5805  & 11610 & 58050 \\
			17    & 0.01  & 149   & 744   & 1487  & 7434  & 14868 & 74337 \\
			17    & 0.001 & 195   & 975   & 1949  & 9741  & 19481 & 97405 \\
			18    & 0.1   & 103   & 515   & 1029  & 5144  & 10288 & 51437 \\
			18    & 0.05  & 118   & 587   & 1173  & 5862  & 11724 & 58620 \\
			18    & 0.01  & 150   & 750   & 1499  & 7491  & 14982 & 74909 \\
			18    & 0.001 & 196   & 980   & 1960  & 9798  & 19596 & 97977 \\
			19    & 0.1   & 104   & 520   & 1040  & 5198  & 10396 & 51976 \\
			19    & 0.05  & 119   & 592   & 1184  & 5916  & 11832 & 59160 \\
			19    & 0.01  & 151   & 755   & 1509  & 7545  & 15090 & 75449 \\
			19    & 0.001 & 198   & 986   & 1971  & 9852  & 19704 & 98518 \\
			20    & 0.1   & 105   & 525   & 1050  & 5249  & 10498 & 52488 \\
			20    & 0.05  & 120   & 597   & 1194  & 5968  & 11935 & 59673 \\
			20    & 0.01  & 152   & 760   & 1520  & 7597  & 15193 & 75962 \\
			20    & 0.001 & 199   & 991   & 1981  & 9904  & 19807 & 99031 \\
			21    & 0.1   & 106   & 530   & 1060  & 5298  & 10595 & 52974 \\
			21    & 0.05  & 121   & 602   & 1204  & 6016  & 12032 & 60160 \\

			\end{array}
			\]\\
			
		\end{center}
	\end{table}%
	\newpage
	\begin{table}[ht]
		\begin{center}
			\[
			\begin{array}{rr|rrrrrr}
			\textbf{K} & {\huge\bm\alpha} & \textbf{ 0.05} & \textbf{ 0.01} & \textbf{ 0.005} & \textbf{ 0.001} & \textbf{ 0.0005} & \textbf{ 0.0001} \\
			\hline
			21    & 0.01  & 153   & 765   & 1529  & 7645  & 15290 & 76450 \\
			21    & 0.001 & 200   & 996   & 1991  & 9952  & 19904 & 99519 \\
			22    & 0.1   & 107   & 535   & 1069  & 5344  & 10688 & 53439 \\
			22    & 0.05  & 122   & 607   & 1213  & 6063  & 12125 & 60625 \\
			22    & 0.01  & 154   & 770   & 1539  & 7692  & 15383 & 76915 \\
			22    & 0.001 & 200   & 1000  & 2000  & 9999  & 19997 & 99984 \\
			23    & 0.1   & 108   & 539   & 1078  & 5389  & 10777 & 53882 \\
			23    & 0.05  & 123   & 611   & 1222  & 6107  & 12214 & 61069 \\
			23    & 0.01  & 155   & 774   & 1548  & 7736  & 15472 & 77359 \\
			23    & 0.001 & 201   & 1005  & 2009  & 10043 & 20086 & 100428 \\
			24    & 0.1   & 109   & 544   & 1087  & 5431  & 10862 & 54307 \\
			24    & 0.05  & 123   & 615   & 1230  & 6150  & 12299 & 61494 \\
			24    & 0.01  & 156   & 778   & 1556  & 7779  & 15557 & 77785 \\
			24    & 0.001 & 202   & 1009  & 2018  & 10086 & 20171 & 100854 \\
			25    & 0.1   & 110   & 548   & 1095  & 5472  & 10943 & 54714 \\
			25    & 0.05  & 124   & 620   & 1239  & 6191  & 12381 & 61901 \\
			25    & 0.01  & 157   & 782   & 1564  & 7820  & 15639 & 78193 \\
			25    & 0.001 & 203   & 1013  & 2026  & 10127 & 20253 & 101262 \\
			26    & 0.1   & 111   & 552   & 1103  & 5511  & 11021 & 55105 \\
			26    & 0.05  & 125   & 623   & 1246  & 6230  & 12459 & 62293 \\
			26    & 0.01  & 158   & 786   & 1572  & 7859  & 15717 & 78585 \\
			26    & 0.001 & 204   & 1017  & 2034  & 10166 & 20331 & 101654 \\
			27    & 0.1   & 111   & 555   & 1110  & 5549  & 11097 & 55482 \\
			27    & 0.05  & 126   & 627   & 1254  & 6267  & 12534 & 62670 \\
			27    & 0.01  & 158   & 790   & 1580  & 7897  & 15793 & 78962 \\
			27    & 0.001 & 205   & 1021  & 2041  & 10204 & 20407 & 102032 \\
			28    & 0.1   & 112   & 559   & 1117  & 5585  & 11169 & 55845 \\
			28    & 0.05  & 127   & 631   & 1261  & 6304  & 12607 & 63034 \\
			28    & 0.01  & 159   & 794   & 1587  & 7933  & 15866 & 79326 \\
			28    & 0.001 & 205   & 1024  & 2048  & 10240 & 20479 & 102395 \\
			29    & 0.1   & 113   & 562   & 1124  & 5620  & 11239 & 56195 \\
			29    & 0.05  & 127   & 634   & 1268  & 6339  & 12677 & 63384 \\
			29    & 0.01  & 160   & 797   & 1594  & 7968  & 15936 & 79677 \\
			29    & 0.001 & 206   & 1028  & 2055  & 10275 & 20550 & 102746 \\
			30    & 0.1   & 114   & 566   & 1131  & 5654  & 11307 & 56534 \\
			30    & 0.05  & 128   & 638   & 1275  & 6373  & 12745 & 63723 \\
			30    & 0.01  & 161   & 801   & 1601  & 8002  & 16004 & 80016 \\
			30    & 0.001 & 207   & 1031  & 2062  & 10309 & 20617 & 103085 \\
			\end{array}
			\]\\		
		\end{center}
	\end{table}
	\cleardoublepage
	\section{CONCLUDING REMARKS}
	We have provided tables that include several selected values of
	sample sizes that relate to typical values of relative abundances of
	a variety of marker species.
	
	The reader can notice that Tables~2 and 3 include repeated sample sizes which correspond to distinct combinations of species proportions. This is explained by the fact that when one proportion is much smaller than the other, the sample size is essentially determined by this small proportion. For example, the sixth, the twelfth, and the twenty-first rows of Table~3 display the same sample sizes even though the values of $p_2$ are different. This is exactly because the smallest of the three $p_2$ values ($p_2=0.005$) is still an order of magnitude larger than $p_1=0.0005$. In other words, when one uses a
	third marker whose proportion is essentially higher than the other
	two proportions, it is enough to consider the first two proportions
	only in order to get a good assessment of $n$.
	
	Since the task of multiple species detection with high likelihood within a short geologic time interval is recurrent in biodiversity and biostratigraphic studies, this paper provides the theoretical framework for the computation of the sample size, together with applied tables, when more than one pre-determined species is to be detected in the same sample, given the required confidence level for each species, and {\it
		\textit{a priori}} assumptions of species proportions in the population.
	\section*{Acknowledgements}
	This research was supported by a former URB – Long-Term Development Grant to Ali T. Haidar. The authors would like to thank Jason Moore (University of New Mexico), Bill Haneberg, as well as many anonymous reviewers for providing precious comments to review an early draft of the manuscript. Special thanks are due to G. Villa (University of Parma, Italy) for indentifying the utility of the database of nannofossil count by Persico \textit{et al.} (2012). She kindly recommended the use of this nannofossil database to emphasize the difference between the sample sizes given by the binomial and those by the multinomial distributions. Mr. Ali Khiyami helped with merging the separate parts of the manuscript into a single latex file, and with some editing of the manuscript.
	\section*{References}
	Agterberg, F.P., 1990. \textit{Automated Stratigraphic Correlation}. Developments in Palaeontology and Stratigraphy, 13. Elsevier. Amsterdam. 424 p.
	\\\\
	Backman, J., Raffi, I., Rio, D., Fornaciari, E., and Pälike, H., 2012. Biozonation and biochronology of Miocene through Pleistocene calcareous nannofossils from low and middle latitudes.\textit{ Newsletters on Stratigraphy} 45(3), pp. 221-244.
	\\\\
	Dennison, J.M., and Hay, W.W., 1967. Estimating the needed sampling area for subaquatic ecological studies. \textit{Journal of Paleontology} 41(3), pp. 706-708.
	\\\\
	Degroot, M.H., and Schervish, M.J., 2012. \textit{Probability and Statistics}. Addison-Wesley, Amsterdam, The Netherlands, 4th edition. 912 p.
	\\\\
	Fatela, F., and Taborda, R., 2002. Confidence limits of species proportions in microfossil assemblages. \textit{Marine Micropaleontology} 45, pp. 174-196.
	\\\\
	Guinasso Jr., N.L., and Schink, D.R., 1975. Quantitative estimates of biological mixing rates in abyssal sediments. \textit{Journal of Geophysical Research} 80, pp. 3032-3043.
	\\\\
	Haidar, A.T., 2015. Entropy Estimate of Superpositional Stratigraphic Time for Shuffled Sediment. \textit{Hannon} 27, pp. 7-34.
	\\\\
	Hayek, L.C., and Buzas, M.A., 1997.\textit{ Surveying Natural Populations. Quantitative Tools for Assessing Biodiversity}. Columbia University Press, New York, USA, 2nd edition, 590 p.
	\\\\
	Hills, S.J., and Thierstein, H., 1989. Plio-Pleistocene Calcareous Plankton Biochronology. \textit{Marine Micropaleontology} 14, pp. 7-96.
	\\\\
	Moore, J.R., Norman, D.B., and Upchurch, P., 2007. Assessing relative abundances in fossil assemblages \textit{Palaeogeography, Palaeoclimatology, Palaeoecology} 253(3–4), pp. 317-322.
	\\\\
	Persico, D., Fioroni, C., and Villa, G., 2012. A refined calcareous nannofossil biostratigraphy for the middle Eocene-early Oligocene Southern Ocean ODP sites. \textit{Palaeogeography, Palaeoclimatology,\\ Palaeoecology} 335-336, pp. 12-23.
	
\end{document}